\documentclass[]{spie}  

\usepackage{amsmath,amsfonts,amssymb}
\usepackage{graphicx}
\usepackage[colorlinks=true, allcolors=blue]{hyperref}

\title{New 50-m-class single-dish telescope: Large Submillimeter Telescope (LST)}

\author[a]{Ryohei Kawabe}
\author[b,c]{Kotaro Kohno}
\author[b]{Yoichi Tamura}
\author[a]{Tatsuya Takekoshi}
\author[a]{Tai Oshima}
\author[b]{Shun Ishii}
\author{LST working group}
\affil[a]{National Astronomical Observatory of Japan, 2-21-1 Osawa, Mitaka, Tokyo, 181-8588 Japan}
\affil[b]{Institute of Astronomy, Graduate School of Science, The University of Tokyo, 2-21-1 Osawa, Mitaka, Tokyo, 181-0015 Japan}
\affil[c]{Research Center for the Early Universe, Graduate School of Science, The University of Tokyo, Hongo, 7-3-1 Bunkyo-ku, Tokyo, 113-0033 Japan}

\authorinfo{Further author information: (Send correspondence to R.K.)\\R.K.: E-mail: ryo.kawabe@nao.ac.jp\\  K.K.: E-mail: kkohno@ioa.s.u-tokyo.ac.jp}

\begin{document} 
\maketitle

\begin{abstract}
We report on a plan to construct a 50-m-class single-dish telescope, the Large Submillimeter Telescope (LST). The conceptual design and key science behind the LST are presented, together with its tentative specifications. This telescope is optimized for wide-area imaging and spectroscopic surveys in the  70--420 GHz frequency range, which spans the main atmospheric windows at millimeter and submillimeter wavelengths for good observation sites such as the Atacama Large Millimeter/submillimeter Array (ALMA) site in Chile. We also target observations at higher frequencies of up to 1 THz, using an inner high-precision surface. Active surface control is required in order to correct gravitational and thermal deformations of the surface, and will be useful for correction of the wind-load deformation. The LST will facilitate new discovery spaces such as wide-field imaging with both continuum and spectral lines, along with new developments for time-domain science. Through exploitation of its synergy with ALMA and other telescopes, the LST will contribute to research on a wide range of topics in the fields of astronomy and astrophysics, e.g., astrochemistry, star formation in our Galaxy and galaxies, the evolution of galaxy clusters via the Sunyaev-Zel'dovich (SZ) effect, the search for transients such as $\gamma$-ray burst reverse shocks produced during the epoch of re-ionization, electromagnetic follow up of detected gravitational wave sources, and examination of general relativity in the vicinity of super massive black holes via submillimeter very-long-baseline interferometry (VLBI). 
\end{abstract}

\keywords{Radio Astronomy, Submillimeter Telescope, Wide-Field Imaging, Imaging Spectroscopy}

\section{INTRODUCTION}
\label{sec:intro}  

As a result of the implementation of the Atacama Large Millimeter/submillimeter Array (ALMA), exciting and cutting edge research is being undertaken in millimeter and submillimeter astronomy. ALMA has unprecedented performance and capabilities, e.g., a high angular resolution (down to 10 milli-arcsecond) and high-sensitivity imaging. On the other hand, the millimeter and submillimeter universe unveiled by ALMA is quite limited as regards sky and spectroscopic coverage, i.e., in terms of the three-dimensional (3D) volume of the universe. 

In 2008, approximately, we began discussions on a future large single-dish telescope. Our initial motivation was the planning of a next-generation millimeter and submillimeter telescope that could, to the greatest possible extent, inherit both the large collecting area of the Nobeyama Radio Observatory (NRO) 45-m telescope (Ukita et al. 1994\cite{Ukita1994}) and the submillimeter capabilities of Atacama Submillimeter Telescope Experiment (ASTE) 10-m telescope (Ezawa et al.~2008\cite{Ezawa2008}). 
Through discussions with researchers in the Europe and US, as well as in East Asia, we further developed this plan, primarily by considering future necessary scientific developments and potential new discoveries that could be complemented by new single-dish telescopes. We finally resolved to construct a large single-dish telescope, i.e., a new 50-m class telescope, to be called the ``Large Submillimeter Telescope (LST)''. Table \ref{tab:specifications} lists the basic specifications of the LST.

The LST will facilitate groundbreaking exploration of an extremely large 3D volume of the universe and, also, completely new advancements in time-domain science for millimeter and submillimeter astronomy. One of the major scientific goals is unveiling the large-scale structure of the high-$z$ universe in 3D, along with elucidation of cosmic star formation history. This can be accomplished via wide-area spectroscopic surveys of dusty starbursts based on CO and [CII] 158 $\mu$m lines (and, presumably, [OIII] 88 $\mu$m lines, as demonstrated by the recent ALMA detection of a $z=7.2$ galaxy; Inoue et al.~2016\cite{Inoue2016}), as well as multi-band continuum surveys. The LST is very complementary to ALMA, and can establish a census of various astronomical objects of our interest through wide-field surveys. The LST can also identify intriguing sources for further investigation by ALMA. By exploiting the synergy with ALMA and other survey-oriented existing and near-future missions in the optical to far-infrared ranges, including HSC/PFS on Subaru, TAO, LSST, Euclid, WFIRST, and SPICA, the LST can contribute to a wide range of research in astronomy and astrophysics. 

In this paper, we briefly summarize the key science behind the LST, the requirements for the LST, the tentative telescope design concept, and the key focal plane instruments.

\begin{table}[ht]
\caption{Basic specifications of the Large Submillimeter Telescope.} 
\label{tab:specifications}
\begin{center}       
\begin{tabular}{|l|l|l|l|} 
\hline
\rule[-1ex]{0pt}{3.5ex}  Item & Requirement & Goal & Comment  \\
\hline
\rule[-1ex]{0pt}{3.5ex}  Diameter (D) & 50 m & -  & \\
\hline
\rule[-1ex]{0pt}{3.5ex}  Frequency (f) & 70--420 GHz & 70--420 GHz & Full illumination \\
\rule[-1ex]{0pt}{3.5ex}   & 420--950 GHz  & 420--1200 GHz & Illumination on inner 25--30 m \\
\rule[-1ex]{0pt}{3.5ex}   &  &  & wind $<$10 m/s, PWV $<$0.2 mm, nighttime \\
\hline
\rule[-1ex]{0pt}{3.5ex}  Spatial Resolution ($\theta$) & $22''$ -- $3.6''$ & $22''$ -- $3.6''$ & for 70--420 GHz\\
\hline
\rule[-1ex]{0pt}{3.5ex}  Field of View (FoV) & 0.5 deg & 1.0 deg & diameter (circular FoV) \\
\hline 
\rule[-1ex]{0pt}{3.5ex}  Operation Period & 30 year & -- & \\
\hline 
\rule[-1ex]{0pt}{3.5ex}  Candidate Site & Atacama/Chile & -- & \\
\hline 
\rule[-1ex]{0pt}{3.5ex}  Radome & None & -- & \\
\hline 
\end{tabular}
\end{center}
\end{table}

   \begin{figure} [ht]
   \begin{center}
   \begin{tabular}{c} 
   \includegraphics[width=13cm]{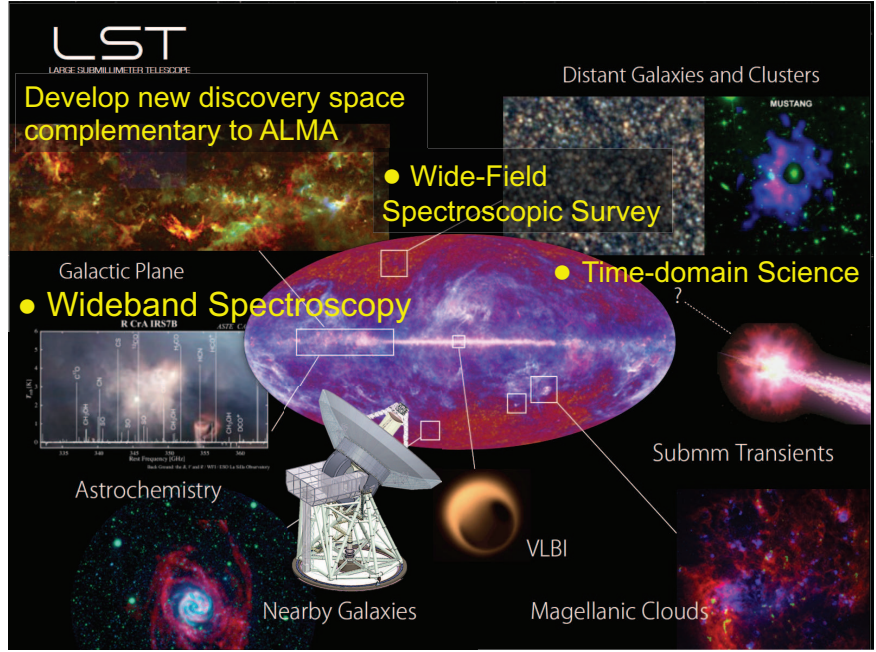}
	\end{tabular}
	\end{center}
   \caption[example] 
   { \label{fig:LST-science} 
Science cases which will be explored by LST.}
   \end{figure}

\section{KEY SCIENCE OF NEW TELESCOPE}

As shown in Figure \ref{fig:LST-science}, the LST will contribute to research on a wide range of topics in the fields of astronomy and astrophysics, e.g., investigations of the chemical evolution from protostellar cores to protoplanetary disks, star formation in our Galaxy and galaxies, the evolution of galaxy clusters via the Sunyaev-Zel'dovich (SZ) effect, the search for submillimeter transients such as $\gamma$-ray burst (GRB) reverse shocks produced at the epoch of re-ionization (EoR) via high cadence wide sky surveys, electromagnetic follow up of detected gravitational wave sources with extremely large positional uncertainty (on a scale of a few 10 deg$^2$ scale, even in the LIGO-VIRGO-KAGURA era), and examination of general relativity in the vicinity of super massive black holes (SMBHs) via submillimeter VLBI. Details of these science cases will be given elsewhere (The LST white paper, Kawabe et al., in prep.). 

In this section, we describes some selected key science cases, which determine the major requirements for the telescope and focal plane instruments. 

\subsection{Exploration of Cosmic Star Formation History and Large-Scale Structures via Multi-band Continuum Survey and CO/[CII] Tomography}
\label{sec:key_science} 

One of the major scientific goals of the new telescope is to determine the complete history of star formation accross cosmic time, by uncovering a statistically large number of dusty star-forming galaxies (e.g., Casey et al.~2014\cite{Casey2014}). Wide-area sky surveys at millimeter and submillimeter wavelengths will play crucial roles in this investigation, because the cosmic infrared background light (CIB) uncovered by COBE indicates that approximately half of the radiation produced by stars and/or accreting SMHBs has been hidden by dust. Millimeter and submillimeter wavelengths are best suited for detecting such re-processed thermal dust emission in the early Universe via the strong negative K-correction. 

Wide-field multi-band continuum imaging surveys with a high angular resolution (a few arcsec) will allow us to uncover the large-scale structures of dusty star-forming galaxies. A resolution of a few arcsec is essential for efficient and reliable multi-wavelength counterpart identification and the resultant determination of the photometric redshifts of these dusty galaxies without interferometric follow-up. Recent high-resolution ALMA studies of submillimeter galaxies uncovered by ground-based 10--15 m class telescopes and Herschel/SPIRE (all with $>10$-arcsec resolutions) reveal that these galaxies are often composed of multiple sources due to a large observing beam (e.g., Simpson et al.~2015\cite{Simpson2015}).

 
  Another novel approach to elucidation of cosmic star formation history is a blind search for CO and [CII] emissions. In particular, searching for [CII] emitters in the appropriate frequency range allows us to sample those sources very efficiently for a redshift range of 3.5 to 9 (190 to 420 GHz), reaching the star-formation in the EoR. Further, spectroscopic analysis of CO in the lower frequency bands offers an opportunity to constrain the evolution of CO luminosity functions across cosmic time. 
This approach is refereed to as ``CO/[CII] tomography'', because it is a CO/[CII] analog to the HI 21-cm tomography which will be conducted with the Square Kilometer Array (SKA). A feasibility study of CO/[CII] tomography has been conducted based on a mock galaxy catalog containing 1.4 million objects with CO flux $S_{\rm CO}\delta v \geq 0.01$ Jy km s$^{-1}$, drawn from the S$^3$-SAX project (Obreschkow et al.~2009\cite{Obreschkow2009}). 

\begin{figure} [ht]
   \begin{center}
   \begin{tabular}{c} 
   \includegraphics[width=17cm]{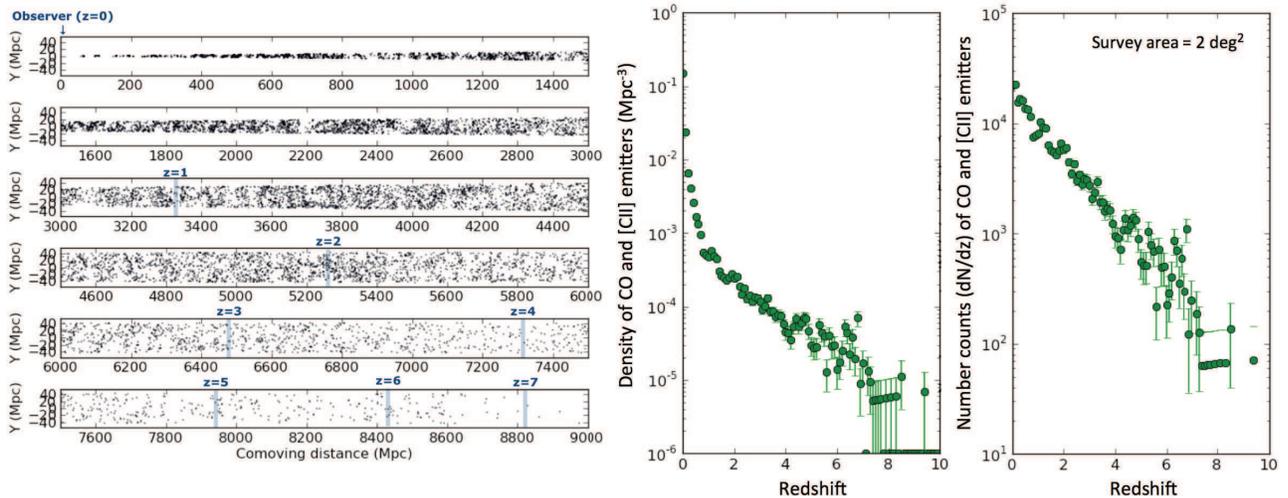}
	\end{tabular}
	\end{center}
   \caption[example] 
   { \label{fig:tomography} 
A feasibility study of CO/[CII] tomography using LST equipped with a multi-pixel imaging spectrometer instrument (Tamura et al., in prep.). (Left) A simulated light cone of the proposed 2 deg$^2$ spectroscopic deep survey. (Middle) The expected volume density of CO and [CII] emitting galaxies as a function of redshift. (Right) Number counts of CO and [CII] emitters.}
   \end{figure}

Figure \ref{fig:tomography} demonstrates that a blind spectroscopic galaxy survey using a 50-m telescope equipped with a 100-pixel dual-polarization receiver array, which covers the 70--370 GHz wavebands simultaneously, will be promising indeed. The proposed reference survey of 2 deg$^2$ in 1,000 hr (on-source) will uncover $> 10^5$ line-emitting galaxies ($\geq 5\sigma$) in total, including $\sim 10^3$ [CII] emitting galaxies in the EoR (Tamura et al., in prep.). 
  %
  Such a spectroscopic deep survey will not be severely affected by source confusion noise, unlike continuum surveys, and can potentially detect fainter high-$z$ galaxies. Recent successful blind or serendipitous detections of CO- and [CII]-line-emitting galaxies using ALMA (e.g., Tamura et al.~2014\cite{Tamura2014}) also advance our plan.
  
  Further important scientific goals are the characterization of dark halo masses hosting these star-forming populations, which are traced by CO and [CII] lines, and study of the growth rate of cosmic large-scale structures by measuring redshift space distortion (RSD) for CO-emitting galaxies at a redshift range beyond 2. These data  will constitute complementary RSD measurements to those obtained via optical/near-infrared redshift surveys (e.g., Okumura et al.~2016\cite{Okumura2016} and references therein). Approximately $10^4$ spec-$z$ galaxies will be required for RSD measurements, and a survey area  that is 5$\times$ wider than that shown in Figure \ref{fig:tomography} (i.e., 5,000 hr survey in total) will be sufficient for cosmological purposes (10 deg$^2$) and a large number of galaxies up to $z\sim3$. 


\section{SITE CONDITIONS AND TELESCOPE REQUIREMENTS}

\subsection{Site and Operation Conditions}

Before determining the detailed requirements for the submillimeter telescope, we investigated the weather statistics of a possible site for the LST, i.e., the ALMA site. We found that, for a fraction of time in the year (except for two months in the Altiplano winter, i.e., January and February) there is an approximately $50 \%$ probability of a wind speed of  $\leq  10$ m/s and precipitable water vapor (PWV) $\leq 4$ mm. We set these conditions as the basic operation conditions for the submillimeter observations at frequencies higher than 300 GHz.  We expanded the operation condition to a wind speed range of $\leq 20$ m/s for millimeter observations. Note that the basic LST requirements described below were developed for the submillimeter operation conditions.

\subsection{Diameter and Surface Accuracy}

The large diameter of the future submillimeter telescopes is key to its cosmological research applications as discussed in section \ref{sec:key_science}. The large diameter facilitates a large collecting area (i.e., high sensitivity), less confusion noise, and a small beam size; hence, deeper surveys and reliable identification of multiwavelength counter parts to uncovered subumillimeter sources will be facilitated. If we realize a 50-m-class submillimeter single-dish telescope, we can improve the point source sensitivity by a factor of $\sim20$, the confusion noise by a factor of $\sim10$, and the beam size by $\sim5$ compared with the existing submillimeter dishes in Chile, such as ASTE 10-m and Atacama Pathfinder Experiment (APEX) 12-m telescopes.

The total surface accuracy (in rms) should be better than 1/16th of the observing wavelength so as to achieve high aperture and beam efficiencies. We take 420 GHz (714 $\mu$m) as being the highest observing frequency for full aperture use; hence, the surface accuracy is required to be $\leq 45$ $\mu$m. To achieve this accuracy under the submillimeter operation conditions, active surface control is essential. Note that the current LST requirements do not include an astrodome.

\subsection{Wide Field-of-View}

In order to conduct extremely large-area ($>$a few 100 -- 1000 deg$^2$) cosmological deep surveys and high-cadence surveys for transient sources, a wide FoV of 0.5 deg$^2$ (up to 1 deg$^2$) is required. The FoV will be shared by large-format cameras, imaging spectrometers, and large-format heterodyne receiver arrays. The receiver cabin should have sufficient room for these instruments and related warm optics. 

\subsection{Observing Frequency and 1-THz Challenge}

ALMA is currently in operation at a frequency range of 80--950 GHz at the Atacama site in Chile. To maximize synergy with ALMA, the LST observations must also be conducted in the same frequency range. 
However, observations at submillimeter windows, e.g., 690 GHz (450 $\mu$m) and 850 GHz (350 $\mu$m) are quite tough, because the availability of the excellent atmospheric conditions suited for such short-wavelength observation is limited, even at the Atacama site. If the LST has a significantly large collecting area, it can effectively exploit ``the submillimeter weather''. The necessity to employ the full aperture of the LST for  observations at such high frequencies has quite a significant impact on the telescope design and manufacture, especially as regards the surface and pointing accuracies. Our current plan for the $\geq 420$-GHz frequency range is to use the central high-precision area ($\leq 25$ $\mu$m in rms), which will be $\sim 30$ m in diameter, and weather that is superior to the submillimeter conditions, i.e., a low PWV (e.g., $< 0.5$ mm) and less-windy conditions.            

\subsection{Pointing Accuracy}

The current specification for pointing under the submillimeter observation conditions is $\leq 0.7 $ arcsec, which corresponds to a gain degradation of $\sim10\%$ at both 420 (full aperture; FWHM of beam $\sim4$ arcsec) and 690 GHz (30-m aperture; also $\sim 4$ arcsec). Our pointing accuracy goal is $\sim 10\%$ of the FWHM, i.e., $\sim0.4$ arcsec, which will be a challenge.

\section{TELESCOPE DESIGN CONCEPT}
\label{sec:telescope_design}

This section presents the outcomes of the preliminary design study for the telescope. 

\subsection{Existing Telescopes and Guidance for LST}

Millimeter telescopes such as the NRO 45-m and IRAM 30-m telescopes are very successful. In these telescopes, homology deformation design is applied to mitigate performance degradation due to gravitational deformation. Thermal control of the back-up structure (BUS) and telescope mount is also incorporated (Baars 2007\cite{Baars2007}, Greve \& Bremer 2010\cite{Greve2010}).

Radio holography measurements have been developed and used to improve the surface accuracy of these telescopes, although an active surface control system is not employed. However, an active surface control system has been implemented in the Large Millimeter Telescope (LMT; Hughes et al.~2010, 2016\cite{Hughes2010,Hughes2016}) in Mexico, a next generation telescope (Schloerb et al.~2016\cite{Schloerb2016}). 

The ALMA 12-m and 7-m submillimeter antennas constitute a breakthrough in terms of telescope performance, having excellent surface accuracy ($\sim20$ $\mu$m) and pointing accuracy. For these antennas, new technology such as a meteorology system, a carbon-fiber-reinforced plastic (CFRP) BUS, and very high-precision surface panels have been developed and implemented. In addition, operation and maintenance experience at a high-altitude site (5,000 m) has been accumulated. The technology developed for application to a submillimeter telescope, together with that produced for large-millimeter single-dishes, provides guidance for the future design of large submillimeter telescopes with $\sim50$-m diameter.  

\subsection{Active Surface Control and Adjustable Small-Size Surface Panel}

A key technology for the LST is an active surface control system. Small adjustable surface panels are required to accurately compensate for gravitational and thermal deformation and to achieve high surface accuracy for each panel. A small panel size is also required from the perspective of the machining capability; a larger panel cannot be machined with high precision. The maximal size is $\sim2-3$ m, and several actuators per panel are required for panel adjustment and panel deformation correction.   

\subsection{Ritchey-Chr\'etien Telescope Design}

The optical system for the LST requires a large FoV ($\sim1^\circ$) for survey capability with a large-format ($>$ mega-pixel) camera system using direct detectors. On the other hand, the collecting area and spatial resolution of the 50-m telescope at submillimeter wavelength are also powerful tools for single pointing or smaller (a few arcminutes) FoV instruments (e.g., THz instruments and 1000-pix class heterodyne cameras), which require a Nasmyth cabin for simultaneous operation of several instruments.
In this case, a Cassegrain F-number of 6 is appropriate for realizing a compact transmission system with $1^\circ$ Nasmyth optics.

Aberrations are another limiting factor of the FoV as they decrease the phase coupling efficiency of the optical system. The Ritchey-Chr\'etien (RC) system provides a wider FoV, which is limited by the astigmatism aberration. This is in comparison to the classical design, which is limited by comatic aberration.

\begin{figure} [ht]
 \begin{minipage}{0.5\hsize}
  \begin{center}
   \includegraphics[width=75mm]{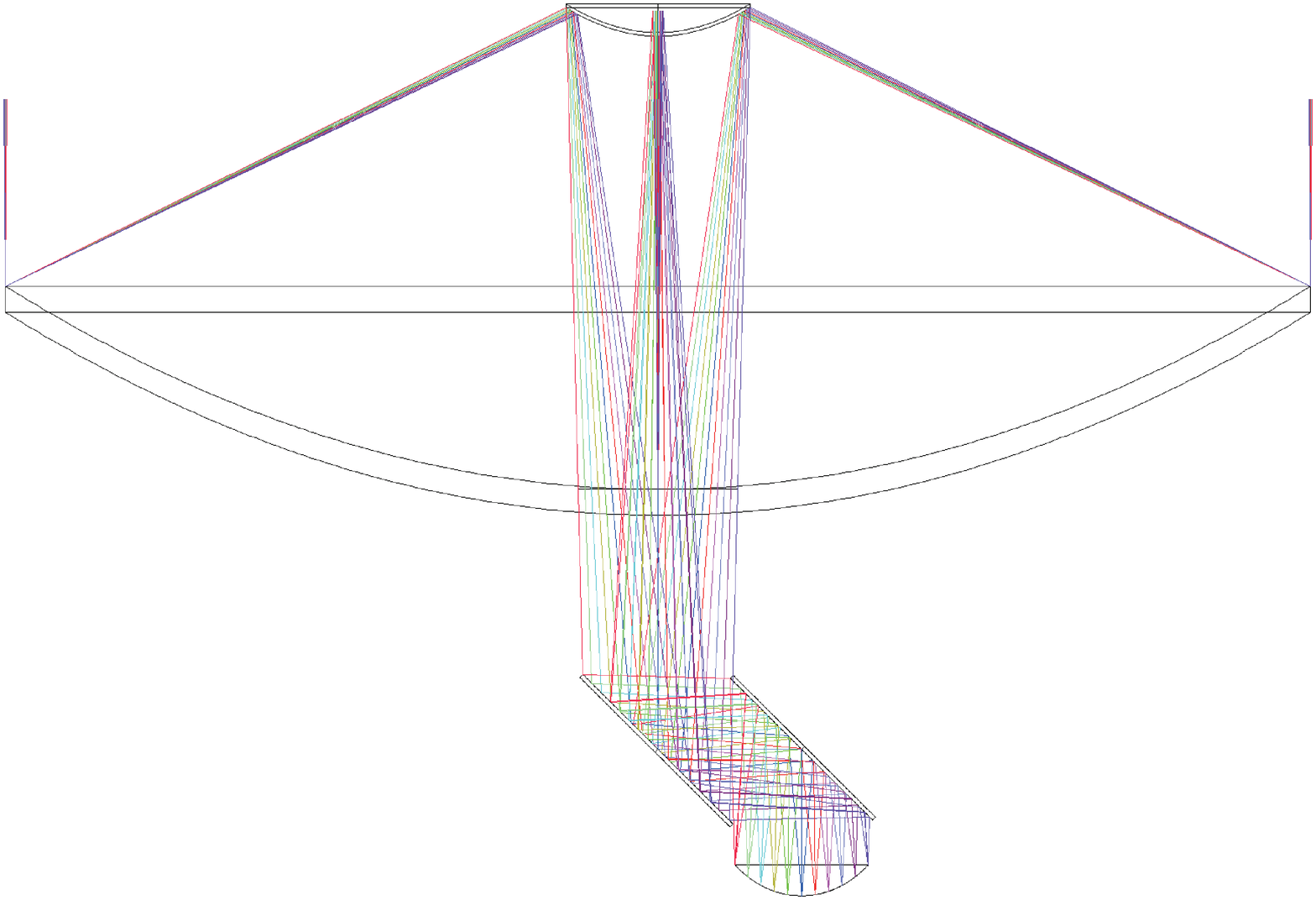}
  \end{center}
  \caption{A practical Ritchey-Chr\'etien F/6 design of the LST optics. The telescope diameter is 50 m, and the diameter of the secondary mirror is 6.6 m. The detailed optics parameters are given in Table \ref{table:cassegrainParameters}.}
  \label{fig:optics-trace}
 \end{minipage}
   \begin{minipage}{0.1\hsize}
  \end{minipage}
 \begin{minipage}{0.5\hsize}
  \begin{center}
   \includegraphics[width=80mm]{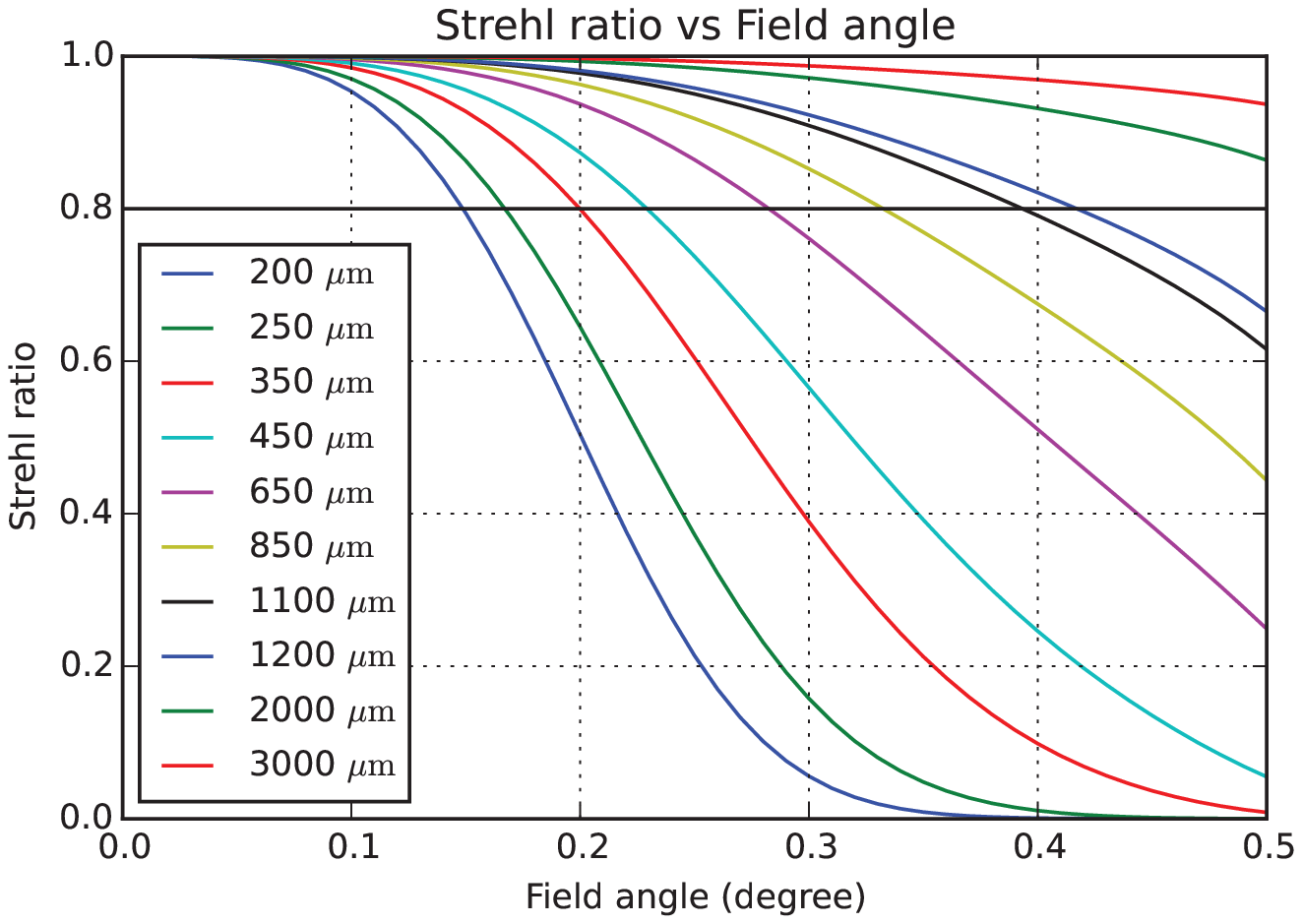}
  \end{center}
  \caption{Strehl ratios as a function of Field angle, displaying the achievable field of view based on the current optics design of LST.}
  \label{fig:Strehl_ratios}
 \end{minipage}
   \end{figure}

Figure \ref{fig:optics-trace} shows the practical RC F/6 design of the LST optics. The Cassegrain parameters are given in Table \ref{cassegrainParameters}. The large sub-reflector is essential for improvement of the aberrations, and to ensure sufficient distance between the secondary reflector and the focal plane for the Nasmyth optics. The tertiary and quaternary plane mirrors allow the ray to enter the Nasmyth cabin, and can also be used for aberration corrector mirrors to improve the aberration via high-order deformation.


Figure \ref{fig:Strehl_ratios} shows the Strehl ratio of the designed optics without correction of the plane mirrors. For the 2- and 3-mm bands, the observable FoV of $1^\circ$ is limited by the telescope structure and mirror size. At the 850-$\mu$m band, it is possible to obtain an FoV of $0.66^\circ$ with a Strehl ratio of more than 0.8. Under higher-order correction of the plane mirrors, however, a full FoV of $1^\circ$ can be obtained at the main observing frequency of the LST.

The Cassegrain focal plane diameter of 5.1~m is not realistic for filling of the single camera cryostat. Therefore, it is necessary to divide this diameter by the smaller FoV at the focal plane. The maximum size of the cryostat window or lens can be up to 1~m; therefore, the $12'$ diameter focal plane can be allocated to the 19 cryostats. A possible alternative is to use a cryogenically cooled dielectric lens to reduce the focal plane diameter; such plans are also being investigated.

Hereafter the specific optics design of the receiver system will be discussed in conjunction with the instrumental designs.

   
\begin{table}
  \caption{The Cassegrain parameters \label{cassegrainParameters}}
  \label{table:cassegrainParameters}
  \begin{center}
    \begin{tabular}{|l|c|}
      \hline
      \multicolumn{2}{|l|}{\bf{Cassegrain parameters}} \\
      \hline
      Primary diameter & 50,000~mm \\
      Secondary diameter & 6,600~mm \\
      Primary F-number & 0.4 \\
      Cassegrain F-number & 6.0 \\
      \hline
      \multicolumn{2}{|l|}{\bf{Reflectors}} \\
      \hline
      Primary radius of curvature & 40,000~mm \\
      Primary Conic constant & -1.00131137 \\
      Secondary radius of curvature & -5509.821~mm \\
      Secondary Conic constant & -1.31866832 \\
      Primary to secondary distance & 17,428.75~mm \\
      Primary to Cassegrain focus & 21,140~mm \\
      Secondary to Cassegrain focus distance & 38,568.75~mm \\
      Actual secondary diameter & 7,046.152~mm \\
      Vertex hole diameter & 6,109.115~mm \\
      \hline
      \multicolumn{2}{|l|}{\bf{Focal plane}} \\
      \hline
      Diameter & 5,115.485~mm \\
      Curvature & -3,141.814~mm \\
      Conic constant & -0.307066 \\
      \hline
    \end{tabular}
  \end{center}
\end{table}

\subsection{Preliminary Conceptual Design}

\begin{figure} [ht]
 \begin{minipage}{0.5\hsize}
  \begin{center}
   \includegraphics[width=70mm]{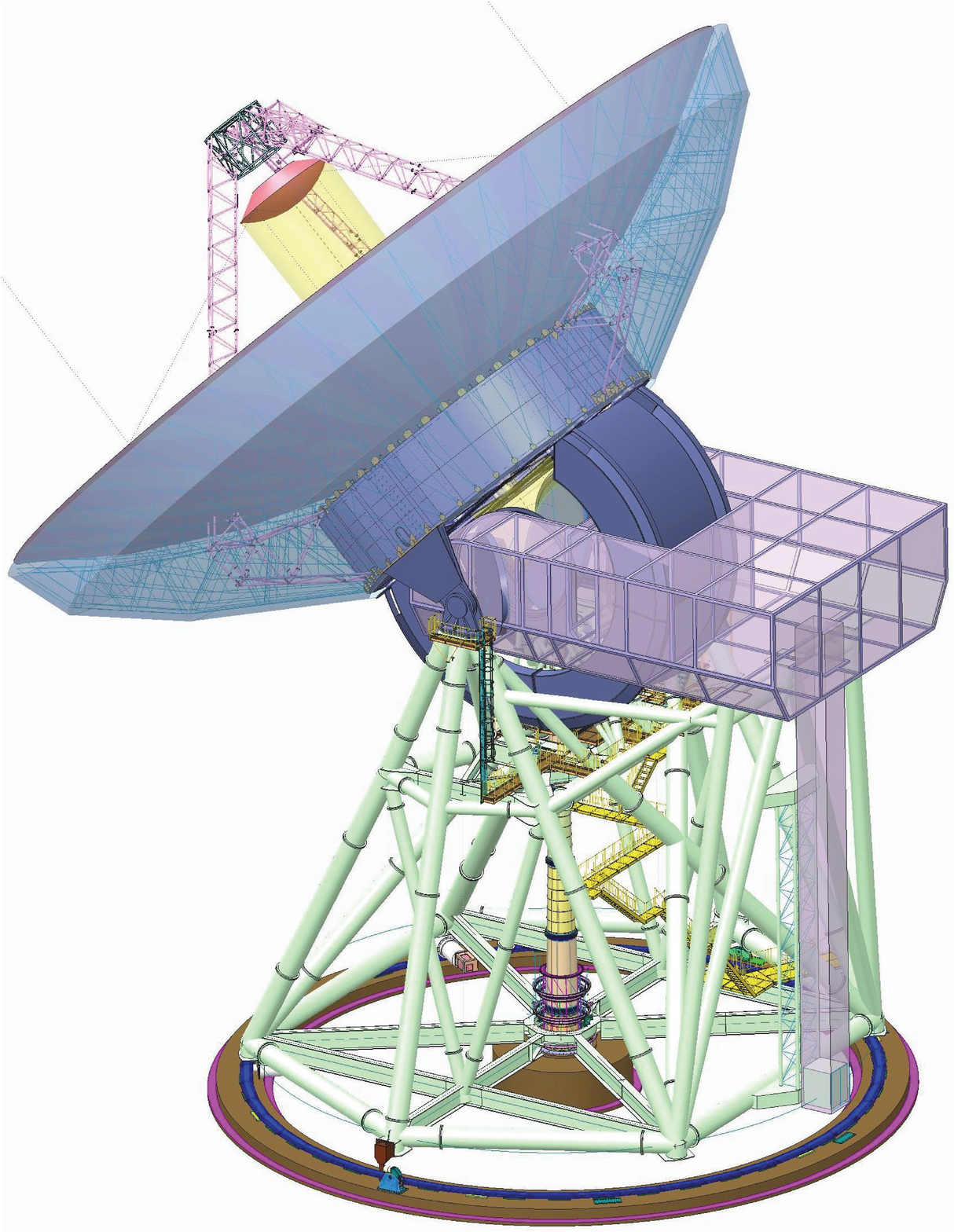}
  \end{center}
  \caption{A conceptual figure of the LST. Courtesy of Mitsubishi Electric Corporation (MELCO).}
  \label{fig:telescope_side_view}
 \end{minipage}
  \begin{minipage}{0.1\hsize}
  \end{minipage}
\begin{minipage}{0.46\hsize}
  \begin{center}
   \includegraphics[width=75mm]{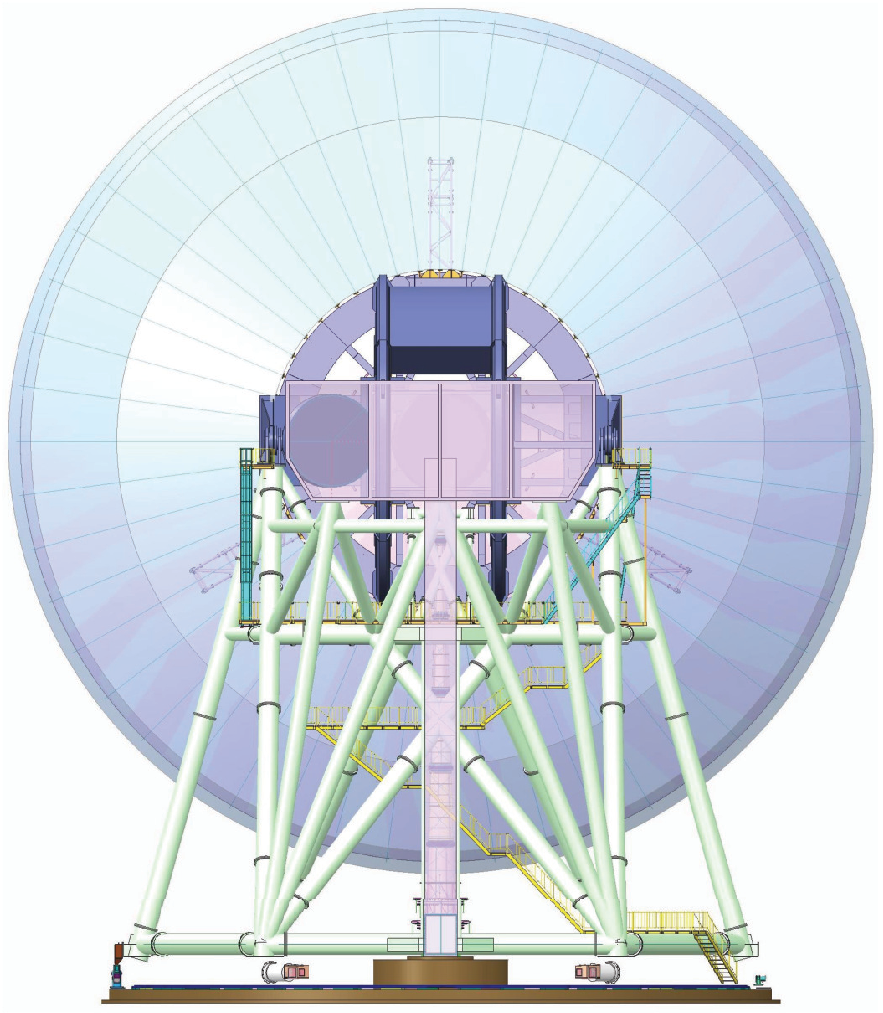}
  \end{center}
  \caption{A rear-side view of the LST. Courtesy of MELCO. }
  \label{fig:telescope_rear_view}
 \end{minipage}
   \end{figure}

\begin{figure} [ht]
 \begin{minipage}{0.46\hsize}
  \begin{center}
   \includegraphics[width=70mm]{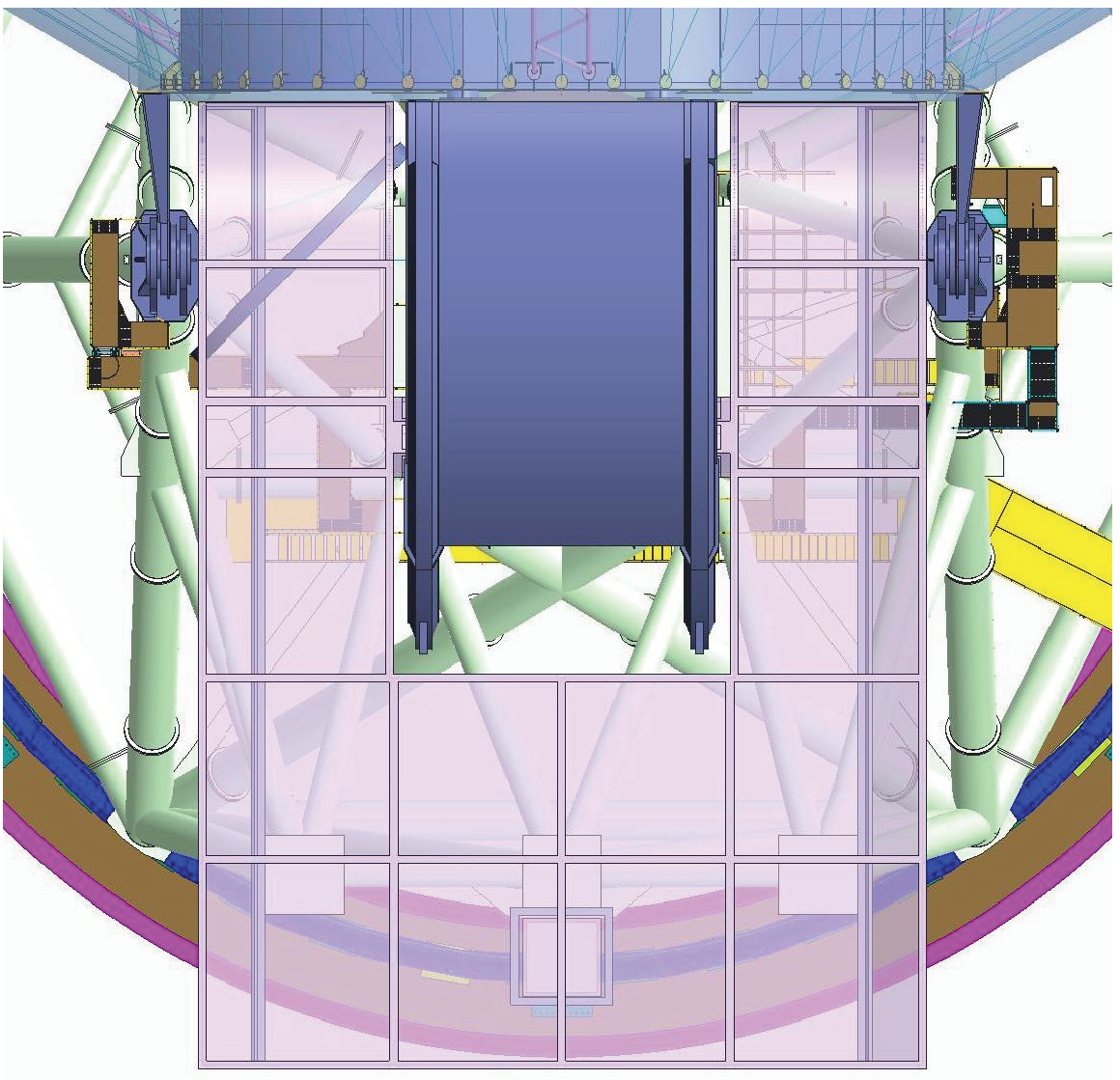}
  \end{center}
  \caption{A top view of LST showing the close-up view of the receiver cabin and the  double sector gear. Courtesy of MELCO.}
  \label{fig:RXcabin}
 \end{minipage}
   \begin{minipage}{0.1\hsize}
  \end{minipage}
 \begin{minipage}{0.5\hsize}
  \begin{center}
   \includegraphics[width=70mm]{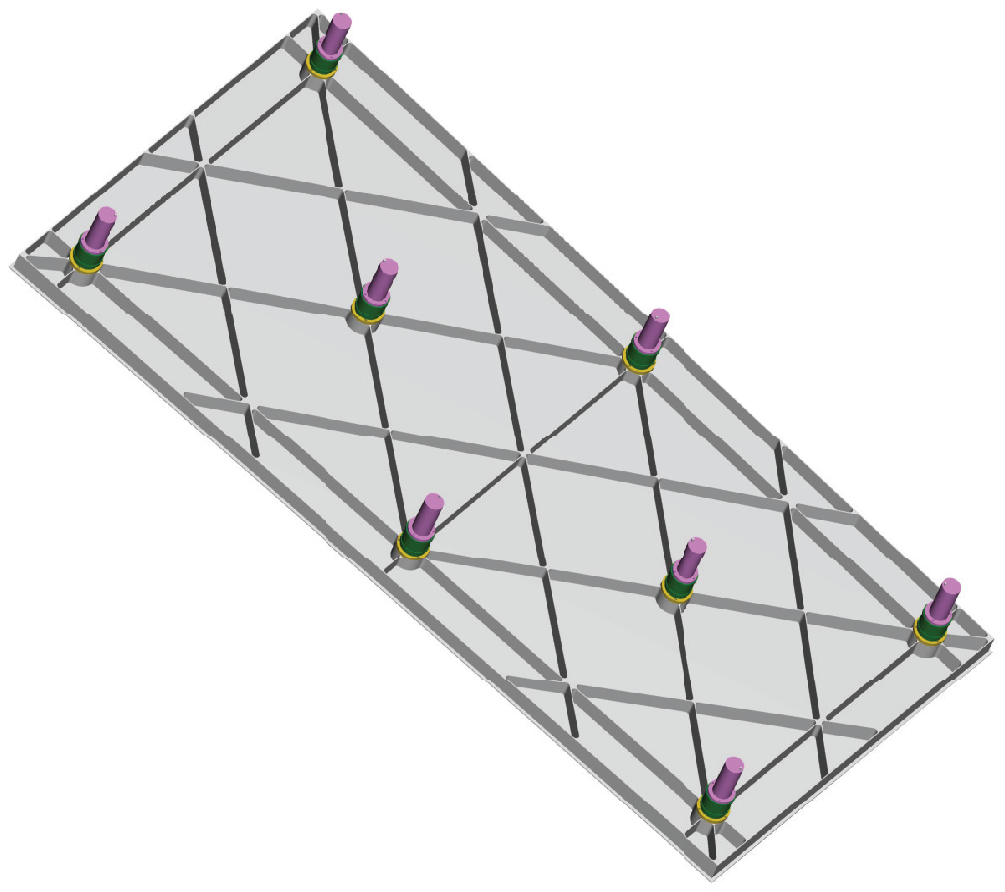}
  \end{center}
  \caption{A rear-side view of a surface panel with limb structures and actuators. The dimension is 2 m $\times$ 1 m. Courtesy of MELCO. }
  \label{fig:panel}
 \end{minipage}
   \end{figure}

 A sample conceptual designs for the LST has been developed and tentative figures of the telescope design have been provided by one of the antenna manufacturers (see Figures \ref{fig:telescope_side_view}--\ref{fig:panel}). Some major requirements for the LST (e.g., tentative wide-FoV optics and a large receiver cabin) have been accommodated in this mechanical design, which reveals the LST appearance and the influences of the various specifications on the mechanical design. However, quantitative investigations of the LST performance specifications, such as the surface and pointing accuracies, are not fully included in this design study, and are important items for future investigation. Note that the tentative surface error budget for the LST is also shown in Table \ref{tab:Error_Budget}. 

\begin{table}[ht]
\caption{Tentative surface error budget for LST and comparison with IRAM 30-m telescope (Baars et al.~1987\cite{Baars1987}). All unit is $\mu$m in rms.} 
\label{tab:Error_Budget}
\begin{center}       
\begin{tabular}{|l|l|l|l|} 
\hline
\rule[-1ex]{0pt}{3.5ex}  Telescope & IRAM 30-m & LST 50-m & Note \\
\hline
\rule[-1ex]{0pt}{3.5ex}  Gravity (residual) & 40 & 15 & FEM modelling + active surface control \\
\hline 
\rule[-1ex]{0pt}{3.5ex}  Thermal (residual) & 20 & 15 & FEM modelling + active surface control \\
\hline 
\rule[-1ex]{0pt}{3.5ex}  Wind (residual) & 30 & 25 & IRAM spec is for wind velocity $\leq$ 12 m/s \\
\rule[-1ex]{0pt}{3.5ex}                  &    &    & Wind load correction using pressure sensors \\
\hline 
\rule[-1ex]{0pt}{3.5ex}  Surface panel & 26 & 20 & \\
\hline 
\rule[-1ex]{0pt}{3.5ex}  Subreflector (residual) & 20 & 15 & Correction with active surface control\\
\hline 
\rule[-1ex]{0pt}{3.5ex}  \#3, 4 mirrors (residual) & 10 & 15 & Correction with active surface control\\
\hline 
\rule[-1ex]{0pt}{3.5ex}  Measurements and setting errors & 35 & 15 & Holography using astronomical sources\\
\hline 
\rule[-1ex]{0pt}{3.5ex}  Total (RSS) & 70 & 44.1 & \\
\hline
\end{tabular}
\end{center}
\end{table}

\section{FOCAL PLANE INSTRUMENTS}

There are three key instruments for the new telescope, as described in the following. 
Such instruments are increasingly required to have larger FoVs in order to satisfy the camera requirements, along with larger instantaneous bandwidths for heterodyne receivers. In addition, an imaging spectrograph to facilitate ``3D imaging'' of the cosmic large-scale structure has been newly developed. Therefore, prototyping and test of the key instruments are hence strongly required.

\subsection{Ultra-Wideband Medium-Resolution Imaging Spectrometer Array}

One of the key instruments is a multi-object spectrograph and/or imaging spectrometer, which facilitates multi-pixel broadband spectroscopy at the focal plane. Such instrument is strongly desirable for the blank-field CO/[CII] survey. The Deep Spectroscopic High-redshift Mapper (DESHIMA, Endo et al.~2012\cite{Endo2012}) or its scaled-up instrument, tentatively named ``Super-DESHIMA'', as well as the proposed multi-object spectrograph concept, MOSAIC (Baselmans et al.\footnote{\url{http://alma-intweb.mtk.nao.ac.jp/~diono/meetings/ASTE_ALMA_2014/aste-workshop-2014-mosaic.pdf}}), are potential instruments for that purpose. 

DESHIMA is a medium-resolution ($R = \lambda/\delta\lambda \sim 500$) spectrometer with integrated on-chip filterbanks exploiting a microwave kinetic inductance device (MKID). 
A similar concept of on-chip spectrograph has also been proposed (e.g., Kovacs et al.~2012\cite{Kovacs2012}). DESHIMA is designed to span a range of 326--905 GHz, with a number of spatial pixels $N_{\rm pix}$ of up to 7. For our future single-dish telescope, we require the capabilities of ``Super-DESHIMA'', which spans 190--420 GHz (or 70--420 GHz, if possible) with $N_{\rm pix}$ $\geq$ 100--300, in order to conduct the CO/[CII] tomography discussed in section \ref{sec:key_science}.

\subsection{Multiple-Chroic Wide-Field Camera}
Another key instrument is a wide-field, multi-color continuum camera or multiple-chroic camera, covering 150 to 350 GHz with 3 bands (baseline plan). A maximum of 6 bands from 90 to 420 GHz may be achieved (goal).

Large-format multi-color bolometer arrays have already become operational (e.g., SCUBA2; Holland et al.~2013\cite{Holland2013}) employing transition edge sensors (TES) with time-domain multiplexing. TES arrays with frequency-domain multiplexing are now widely implemented, as highlighted by recent scientific results from the South Pole Telescope (Carlstrom et al.~2011\cite{Carlstrom2011}). Multi-chroic TES technology has been intensively studied and developed for cosmic microwave background (CMB) experiments such as Polarbear and LiteBIRD (e.g., Suzuki et al.~2012\cite{Suzuki2012}). 

For the ASTE 10-m telescope, a dual-color TES camera has been developed (e.g., Takekoshi et al.~2012\cite{Takekoshi2012}, Oshima et al.~2013\cite{Oshima2013}, Hirota et al.~2013\cite{Hirota2013}) and commissioning and initial scientific verification were conducted successfully in 2014, at 270 and 350 GHz ($N_{\rm pix}$ = 169 and 271, respectively). Further commissioning runs with a new calibration device are now underway (April--July 2016). An upgrade plan, which will have 881 pixels in total at 350 and 670 GHz, has also been investigated. 

MKID technology is also becoming suitable for multicolor wild-field imaging instruments, as demonstrated by NIKA2 on the IRAM 30-m telescope (Calvo et al.~2016\cite{Calvo2016}). A 600-pixel MKID camera for the Nobeyama 45-m telescope is also under development (Sekimoto et al.~2014\cite{Sekimoto2014}). Verification of both the TES and MKID technologies on ASTE (and other telescopes such as LMT and IRAM) will be one of the very important steps toward developments of the wide-field camera required for the proposed next-generation submillimeter-wave telescope.

\subsection{Multi-band Heterodyne Array Receivers}
Based on the scientific demands for high-spectral-resolution spectroscopy ($R>300,000$ or $dv<1$ km/s, which is currently impossible to achieve with a direct detector spectrograph like DESHIMA), large heterodyne arrays are also unique and indispensable instruments for future large single-dish telescopes; e.g., arrays at 100, 230, 350, and 490 GHz.
An ultra-wideband coherent receiver system is also desirable for some key science cases of the LST, including the unbiased line surveys for exploration of the chemical diversity in protostellar cores, as well as a blind spectroscopic search for absorption lines against the reverse shock of very high-$z$ $\gamma$-ray bursts (Inoue et al.~2007\cite{Inoue2007}), which shall be conducted immediately after the detection (within a few hours) of a submillimeter transient source. 

The considerable expertise has been acquired by the developments of receivers for the NRO 45m, ASTE, and also for the ALMA (e.g., Sekimoto et al.~2008\cite{Sekimoto2008}, Uzawa et al.~2009\cite{Uzawa2009}, Asayama et al.~2014\cite{Asayama2014}), and it can be used for the future developments of the heterodyne instruments.

Although the current intermediate frequency (IF) bandwidth is limited to $\sim$10 GHz, a novel idea to drastically widen the instantaneous bandwidths of the heterodyne instruments has been proposed, in which radio-frequency (RF) domain multiplexing of a series of SIS mixers is employed (Kojima et al.~2015\cite{Kojima2015}).

\acknowledgments 
 
The authors thank all members of the LST working group and contributors to the science cases of the LST. RK, KK, YT, TO, and SI are supported by the Japan Society for Promotion of Science (JSPS) KAKENHI Grant Numbers 25247019 and 15H02073.

\bibliography{report} 
\bibliographystyle{spiebib} 

\end{document}